\documentclass[11pt,a4paper]{article}
\usepackage{amssymb}
\usepackage{amsthm,epsf}
\usepackage{amsfonts}
\usepackage{epsfig}
\usepackage{graphicx}

\begin{document}
\title {\bf  A Markov Chain-Based Numerical Method for Calculating Network
Degree Distributions \footnote {This research is supported in part
by the National Natural Science Foundation of China through grant 70171059
and by Hong Kong Research Grant Council through grants HKUST6089/00E and HKUST6198/01E}}
\date{}
\maketitle

\author{\centerline {\small Dinghua Shi$^{1}$
 , \ Qinghua Chen$^{1, 2}$ \ and \ Liming Liu $^{3,}$\footnote [4] {The corresponding author}}}

\centerline{\small $^1$Department of Mathematics, College of
Science, Shanghai University,} \centerline{\small Shanghai 200436, China }
 \centerline{\small E-mail address: shidh2001@263.net}

\centerline{\small $^2$College of Mathematics and Computer
Science, Fujian Normal University,}
 \centerline{\small Fuzhou 350007, China}
 \centerline{\small E-mail address: qhdchen@yahoo.com.cn}

\centerline{\small $^3$Department of Industrial Engineering and Engineering Management,}
 \centerline{\small Hong Kong University of Science and Technology,}
 \centerline{\small Clear Water Bay, Kowloon, Hong Kong}
 \centerline{\small E-mail address: liulim@ust.hk}

\abstract {This paper establishes a relation between scale-free
networks and Markov chains, and proposes a computation framework
for degree distributions of scale-free networks. We first find
that, under the BA model, the degree evolution of individual nodes
in a scale-free network follows some non-homogeneous Markov chains.
Exploring the special structure of these Markov chains, we are
able to develop an efficient algorithm to compute the degree
distribution numerically. The complexity of our algorithm is
$O(t^2)$, where $t$ is the number of time steps for adding new
nodes. We use three examples to demonstrate the computation
procedure and compare the results with those from the existing
methods.
\bigskip


\bigskip

}

\section{Introduction}
\ \ \ \   Complex networks describe a wide range of practical systems of high
technological, biological, and social importance [1,2].
For example, the Internet, the World Wide Web (WWW), biological cells and
communities of scientists can all be described as complex networks.

Erd\"{o}s and R\'{e}nyi [3] started the early studies of complex
networks as random graphs in 1960. Many years later, Watts and
Strogats [4]'s construction of the small-world network in 1998
represents an interesting development for the study of complex
networks in that it was motivated by observations of real system
behaviors (e.g., Milgram's six-degree connectivity [5]). A common
feature of the random graph and small-world models is that the
degree distribution (the probability of finding a node with $k$
connections) decays exponentially with the number of connections.
However, empirical evidences from the Internet and WWW, among
other complex networks, show a fundamentally different picture,
i.e., the tail of the degree distribution follows a power law.
This led to the introduction in 1999 of scale-free networks by
Albert, Baraba$\acute{a}$si, and Jeong in their pioneering works
[6-8], and the start of a new phase in the study of complex
networks. Recent studies [9-20] are characterized by empirical
observations of scale-free behaviors in various practical systems
and investigations of the formation mechanisms of scale-free
network. A number of important properties in scale-free networks
have been identified, such as the small-world character, the
emergence of hubs, and robustness and frangibility. These
properties show that scale-free networks can play an important
role in the understanding of many complex and important systems.



Two general features can be observed in many real-world networks:
successive additions of new nodes and certain preference in linking to existing nodes.
Albert, Baraba$\acute{a}$si, and Jeong proposed two mechanisms to characterize the
evolution of a scale-free network [7, 8]: the growth mechanism, starting from
$m_0$ nodes, the network grows at a constant speed, i.e., adding
one node at each time step and connecting to $m (m \leq m_0)$ existing nodes;
the preferential attachment, the chance that an existing node
receives a connection from a new node is proportional to the number
of connections it already has. The authors show
that, under these two mechanisms, a network evolves into
a stationary scale-free state. Its degree distribution follows a
power law with the degree exponent $\gamma = 2.9\pm 0.1$ from simulation analysis
and $\gamma = 3$ from the analytical result.
These results are significant for complex networks and the two mechanisms become the first model,
referred to as the BA model,
by which large networks can self-organize into a stationary scale-free state.
Empirical evidences show that in many networks, the
number of edges grows faster than the number of nodes.
This leads to the investigations of $m$-varying BA models, such as Dorogovtsev and Mendes [20].

Our research is mainly motivated by the following observation.
While analytical solutions of the degree distribution for some
simple models, such as the BA model, can usually be obtained, one
has to resort to simulation for the degree distribution when the
mechanisms in model become more complex. This may inhibit the
further development of the theory on complex networks. In this
paper, we propose an alternative approach. We first find that the
degree evolution of a complex network can be characterized by a
sequence of Markov chains. By carefully analyzing the structure of
these Markov chains, we then develop an efficient numerical method
to compute the degree distribution of complex network models. To
show the feasibility and efficiency of our numerical method, we
compute the degree distribution of the basic BA model and two of
its variants.

We organize the paper as follows. In the next section, we review
some of the existing methods for network degree distributions.
We then use Markov chains to capture network dynamics.
Exploring the special structure of the transition matrices of the
Markov chains, we develop an efficient algorithm to compute the
degree distribution asymptotically. We use this algorithm to
compute the exponent of the degree distribution of the BA model.
In Section 3, we compute the
degree distributions of two $m$-varying BA models. We verify our
approach by showing that our numerical results for the BA model
and its variants match very closely to the existing results from
the analytical and simulation approaches. We conclude the paper in
Section 4 by pointing out some future research opportunities.

\section{A Markov chain-based numerical method}

\ \ \ \ 
With the preferential attachment mechanism of the BA model, the probability that node $i$ receives a connection from an up coming new node is proportional to its own degree $k_{i}$ [7], i.e.,
\begin{equation}
\Pi(k_{i}) = \frac{k_{i}}{\sum_{j}k_{j}}.
\end{equation}
Assuming continuity of $k_{i}(t)$ and treating $\Pi(k_{i})$ as its rate of growth,
$k_{i}(t)$ then satisfies the following dynamic equation [7, 8]
\begin{equation}
\frac{\partial k_{i}}{\partial t}= m\Pi(k_{i})=
m\frac{k_{i}}{\sum_{j}k_{j}}= \frac{k_{i}}{2t}.
\end{equation}
Under the initial condition that $k_i(t_i)=m$, the solution of
this equation leads to
\begin{equation}
k_{i}(t)= m(\frac{t}{t_{i}})^{\beta} \ \ , \ \ \ \beta=\frac{1}{2}
\end{equation}
where $t_i$ is the time when node $i$ joins the network, and the
degree distribution
\begin{equation}
P(k) \sim \ {2m^{2}}k^{-\gamma} \ \ \ , \ \ \ \ \gamma= 3.
\end{equation}
Here, $\beta$ is called the dynamic exponent while $\gamma$ the degree exponent.

The above simple analytical method is often refereed to as the
continuum (mean field) theory. Similar power law results for the
degree distribution are also obtained using different analytical
methods by other authors. For example, with the master-equation approach [14],
Dorogovtsev, Mendes and Samukhin treat the degree $k_{i}(t)$ of a
node $i$ at a fixed time $t$ as a random variable. Thus its
probability $P(k,t_{i},t)$ for the BA model has the following
relation:
\begin{equation}
P(k,t_{i},t+1)= \frac{k-1}{2t}P(k-1,t_{i},t) +
(1-\frac{k}{2t})P(k,t_{i},t).
\end{equation}
Let
\begin{equation}
P(k,t) =\frac{\sum_{t_{i}}P(k,t_{i},t)}{t}.
\end{equation}
Assuming that the limit $P(k) = \lim_{t\to\infty}P(k,t)$ exists
and $\lim_{t\to\infty}t[P(k,t+1)-P(k,t)] = 0$ (this is an additional
condition), the degree distribution satisfies equation:
\begin{equation}
2P(k) - 2\delta_{k,m} = (k-1)P(k-1) - kP(k),
\end{equation}
and the network degree distribution can be obtained as
\begin{equation} P(k) = \frac{2m(m+1)}{k(k+1)(k+2)}. \end{equation}
Krapivsky, Redner and Leyvraz's rate-equation approach [15] focuses on the number $N_{k}(t)$ of nodes with $k$ edges at time $t$. For the BA model, $N_{k}(t)$ is shown to satisfy
\begin{equation} \frac{dN_{k}}{dt} = m\frac{(k-1)N_{k-1}(t)-kN_{k}(t)}{\sum_{k}kN_{k}(t)} + \delta_{k,m}. \end{equation}
Asymptotically, $N_{k}(t) =tP(k)$ and $\sum_{k}kN_{k}(t) = 2mt$,
leading to equation (7).

While the above methods handle simple models, such as the BA model, well,
they do not, so far from the best of our knowledge, render
analytical solutions for more complicated models.
In this case, one can usually use simulation.
While simulation is widely applicable,
it is usually quite time consuming and may not be flexible enough
for in-depth analysis of network behaviors.
Here, we propose
a different approach to capture the network dynamics.

Consider the degree $K_{i}(t)$ of node $i$ at time $t$. Following
the increase of $t$, the sequence  $\{K_{i}(t), t=i,i+1,...\}$ is,
based on the preferential attachment mechanism of the BA model, a
stochastic process with the state space $\Omega=\{m,m+1,...\}$.
Here and below, we use the upper case $K$ to emphasize the fact
that the degree sequence is a stochastic process. The attachment
mechanism also indicates that the future evolution of the process
is independent of the past history, given its current state; but
it is time-dependent. This shows that the process $\{K_i(t)\}$ is
in fact a non-homogeneous Markov chain [21], with time-dependent
transition probability
\begin{equation}
p_{kj}(t+1) = P\{K_{i}(t+1)=j \mid K_{i}(t)=k\}
 = \left\{
 \begin{array}{ll}
 1-\frac{k}{2t}, & j=k \\
 \frac{k}{2t}, & j=k+1 \\
 0, & otherwise
 \end{array}
 \right.
\end{equation}
for $k=m,...,m+t-i$, and
\begin{equation}
p_{kj}(t+1)
 = \left\{
 \begin{array}{ll}
 1, & j=k \\
 0, & j\neq k
 \end{array}
 \right.
\end{equation}
for $k > m+t-i$. 
Thus, the dynamics of a node from the time it joins the network
is described by a non-homogeneous Markov chain
and the whole network (excluding the original nodes)
is completely described by $t$ non-homogeneous Markov chains, where
$t$ is the time of the observation.
Let $P_{i}(t+1)$ be the one-step transition probability matrix of node $i$ at time $t$.
We have,  for $t = i,i+1,...$
\begin{equation}
 P_{i}(t+1) = \left[\matrix{1-\frac{m}{2t} & \frac{m}{2t} & \cr
& \cr
 & 1-\frac{m+1}{2t} & \frac{m+1}{2t} & \cr
 &&\ddots & \ddots & \cr
 &&&1-\frac{m+t-i}{2t} & \frac{m+t-i}{2t} & \cr
 & \cr
 &&&&1 & 0  & \cr
 &&&&&\ddots & \ddots &
 }  \right] .
\end{equation}
Let $f_{i}(t)$ be the probability vector (distribution) of
$K_{i}(t)$ for a given $t$, and
\begin{equation}
F_{t+1}^{(S,T)}= \sum_{i=S}^{T}f_{i}(t+1), ~~~
P(k,t+1)=\frac{F_{t+1}^{(S,t)}(k-m+1)}{t-S+1} .
\end{equation}
Here, $S$ and $T$ are two fixed integers between 1 and $t$. Their
meanings will be clear in the computation procedure later. The
desired degree distribution of the network is then
$P(k)=\lim_{t\to\infty} P(k,t+1)$.


Let us examine (13) to see what is involved in computing the
network degree distribution. It is clear that $ P\{K_{i}(i)=k\} =
1$ if $k=m$ and $0$ otherwise. We then have the initial
probability vector
\begin{equation}\nonumber f_{i}(i) = (1,0,0, ... ) = e_{1}
\end{equation}
for any $i$. By density evolution of Markov chain, the $t+1$-step
probability vector $f_i (t+1)$ is given by
\begin{equation}
f_{i}(t+1)=e_{1}\cdot P_{i}(i+1)\cdot
P_{i}(i+2)\cdot \cdot \cdot P_{i}(t+1) \ ,\ t=i,i+1,...
\end{equation}
where the dots represent matrix multiplications. This, together
with (13), shows that computing the degree distribution requires
the multiplications and summations of an infinite number of
infinite matrices. It is not realistic to expected any meaningful
analytical solution from these computations. Even numerical
computation seems unmanageable. Fortunately, our past experience
in infinite matrix computations [22] with a rectangle-iterative
algorithm guides us to explore the special structure of the
one-step transition matrices. This leads to dramatically
simplified matrix manipulations and a highly efficient algorithm.

We note that while the transition matrices of consecutive nodes are different,
their structural similarities lead to the following relations
\begin{equation} e_1P_i(t)=e_1 P_{1}(t), ~~~~ i=2, 3, ...; ~~t=i+1, i+2, ...
\end{equation}
\begin{equation}
e_1 P_i (t)P_i (t+1) = e_1 P_1 (t)P_1 (t+1), ~~~ i=2, 3, ... ; t=i+1, i+2, ... \end{equation}
and in general
\begin{equation}
e_1 P_i (t)P_i (t+1) \cdot \cdot \cdot P_i (t+s) = e_1 P_1 (t)P_1 (t+1) \cdot \cdot \cdot P_1 (t+s),
\end{equation}
for $i=2, 3, ...;~ t=i+1, i+2, ... $ and $s=2, 3, ... ~$.
Substituting the above relations into
\begin{equation}
F_{t+1}^{(S,T)}= \sum_{i=S}^{T} f_i (t+1) = \sum_{i=S}^{T} e_1 P_i
(i+1) \cdot P_i (i+2) \cdot \cdot \cdot P_i (t+1),
\end{equation}
we obtain the following key relation
\begin{equation}
F_{t+1}^{(S,T)}=((\cdot \cdot \cdot
(e_{1}P_{S}(S+1)+e_{1})P_{S}(S+2)+\cdot \cdot \cdot
)+e_{1})P_{S}(T+1) \cdot \cdot \cdot P_S (t+1).
\end{equation}

The computation of $F_{t+1}^{(S,T)}$ becomes very easy with (20).
We start from the inner most bracket. After one multiplication and
one summation, we obtain a row vector whose first two elements are
nonzero. The second round of multiplication and summation lead to
a row vector with the first three elements being nonzero, and so
on so forth. The final result is a row vector with the first
$(t-S+1)$ elements being nonzero. An efficient algorithm can be
developed to implement this procedure. Obviously, the complexity
of the algorithm is $O(t^2)$.

We plot the $\log-\log$ curves for $P(k, t)$ for some different
$m$ and $t$ as shown in Figure 1, and use the least square method
to fit the exponent $\gamma$ and the coefficient $c$ of the
power-law under the BA model. Table 1 lists the numerical results
for different $m$ and $t$ values. We observe that the degree
exponent is independent of $m$ and the value matches those of
simulation and the analytical solution with the mean field method.
The coefficient of degree distribution $c$ is between $2m^2$ and
$2m(m+1)$, again matching the theoretical value from the mean
field method. Furthermore, results for $m=3$ show that the
coefficient $c$ is independent of $t$, i.e., the network is
stationary.

\begin{figure}[htbp]
  \begin{center}
  \includegraphics[width=8cm]{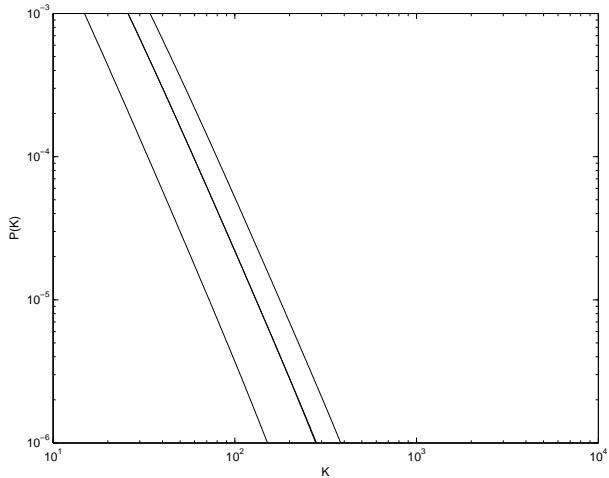}\\
  \caption{{\small The
degree distribution of the BA model} }\label{fig 1}
  \end{center}
\end{figure}

In Figure \ref{fig 1}, the three lines from left to right
correspond to three cases: (1) $m=1,~t=150,000$; (2) $m=3,
~t=100,000, ~150,000,~ 200,000;$ (3) $m=5,~t=150,000$. The line in
case (2) is the overlap of three lines corresponding to three
different $t$ values. This shows that the distribution is
stationary. The three lines of the three cases are parallel, which
further shows the degree exponent of the BA model is independent
of $m$.



\begin{table} [htbp]

 \begin{center}
 \caption{ \ \  Degree exponent and coefficients of the BA model}
 \vspace{3pt}
 \begin{tabular} {|c|c|c|c|}
 \hline
 parameter $m$ & time $t$ & exponent $\gamma$ & coefficient $c$ \\
 \hline
 1 & 150000 & 2.960830 & 3.147515 \\
 3 & 100000 & 2.989636 & 21.79266 \\
 3 & 150000 & 2.990032 & 21.89667 \\
 3 & 200000 & 2.980275 & 21.01711 \\
 5 & 150000 & 2.978894 & 52.58430 \\
 \hline
 \end{tabular}
 \end{center}
 \end{table}



\section{The degree distributions of $m$-varying BA models}
\ \ \ \  Our numerical approach is feasible and can be efficiently
applied to more complex models. Since the number of edges grows
faster than the number of nodes in many networks as shown by
empirical evidences, we compute the degree distributions of two
cases of the BA model with $m$-varying functions in this section.

\subsection{Power function}
\ \ \ \  Let the number of new links added in time step $t$ be
$mt^{\theta}$, $0\leq \theta<1$, i.e.,
the new node $t$ will link itself to $m t^\theta$ different nodes already present in
the system.

We note that after $t$ time steps, this case leads to a random
network with $N=t+m_{0}$ nodes and approximately $\int_{0}^{t}m
x^\theta dx$ links. Then, the total degree number of the system at
time $t$ is
\begin{equation}
\sum_{j}k_{j} \approx  2\int_{0}^{t}m  x^\theta dx =
\frac{2m}{\theta+1}t^{\theta+1}.
\end{equation}

Assuming continuity of $k_i(t)$, it then satisfies the following
dynamic equation
\begin{equation}
\frac{\partial k_{i}}{\partial t}= m t^\theta\Pi(k_{i})= m
t^\theta\frac{k_{i}}{\sum_{j}k_{j}}= \frac{(\theta+1)k_{i}}{2t}.
\end{equation}
Under the initial condition is $k_i(t_i)=m t_i^\theta$, where
$t_i$ is the time when node $i$ joins the network, we solve this
equation and obtain
\begin{equation}
k_{i}(t)= m
t_i^{\theta}(\frac{t}{t_{i}})^{\frac{1+\theta}{2}}=mt^{\theta}(\frac{t}{t_i})^\beta
, \ \ \ \beta=\frac{1-\theta}{2}.
\end{equation}
Hence the degree distribution at time $t$
\begin{equation}
P(k,t) \sim \frac{2}{1-\theta} m^{\frac{2}{1-\theta}} t^z
k^{-\gamma}, \ \ \gamma=
\frac{3-\theta}{1-\theta},\\z=\frac{2\theta}{1-\theta}.
\end{equation}
Here, $z$ is called the non-stationary exponent.
We note that this type of $m$-varying function was first discussed in [20].


We now construct the Markov chain for the degree sequence
$\{K_{i}(t), t=i,i+1,...\}$. The state space is
$\Omega=\{m_{i},m_{i}+1,...\}$, where $m_{i}=m[i^\theta]$. At time
$t$, the probability that an existing node $i$ will connect with
the new node is given by
\begin{equation}
 m t^\theta \frac{k_{i}}{\sum_{j}k_{j}} \approx \frac{(\theta+1)k_{i}
} {2t}.
\end{equation}
Hence, the one-step transition probabilities are
\begin{equation}
p_{kj}(t+1) = P\{K_{i}(t+1)=j \mid K_{i}(t)=k\}
 = \left\{
 \begin{array}{ll}
 1-\frac{(\theta+1)k
} {2t}, & j=k \\
 \frac{(\theta+1)k
} {2t}, & j=k+1 \\
 0, & otherwise
 \end{array}
 \right.
\end{equation}
for $k=m_{i},...,m_{i}+t-i$, and
\begin{equation}
p_{kj}(t+1) = \left\{
 \begin{array}{ll}
 1, & j=k \\
 0, & j\neq k
\end{array}
 \right.
\end{equation}
for $k > m_{i}+t-i$. The transition probability matrix is
\begin{equation}
P_{i}(t+1)=\left[\matrix{1-\frac{m_{i} (\theta +1)}{2t} &
\frac{m_{i} (\theta +1)}{2t} & \cr
 & \cr
 & 1-\frac{(m_{i}+1) (\theta+1)}{2t} & \frac{(m_{i}+1) (\theta+1)}{2t} & \cr
 &&\ddots & \ddots & \cr
 & \cr
 &&&1-\frac{(m_{i}+t-i)(\theta+1)}{2t} & \frac{(m_{i}+t-i)(\theta+1)}{2t}& \cr
 & \cr
 &&&&1 & 0  & \cr
 &&&&&\ddots & \ddots &
 }  \right]
\end{equation}
for $t = i,i+1,...$ .

We now provide the computation results when $\theta=0.2$. We note
that the structure of the transition matrices here is similar to
that of (12). The difference is that now $m_i$ is not a constant,
in general, but a step function of $i$, as shown in Table 2.

\begin{table} [htbp]

 \begin{center}
 \caption{ \ \  Intervals of $m_i$ keep constant}
 \vspace{5pt}
 \begin{tabular} {|c|c|c|c|c|c|c|c|c|c|c|}
 \hline
time $i$ & 32 &  243& 1024 &  3125 & 7776 & 16807 & 32768 & 59049 & 100000 & 161051\\
 \hline
$[t^{0.2}]$ & 2& 3 & 4 & 5 & 6 & 7 & 8 & 9 & 10 & 11  \\
 \hline
 \end{tabular}
 \end{center}
 \end{table}

Therefore, relations (16), (17) and (18) hold for each interval,
e.g., the interval $(243,1023)$. 
Thus we obtain the following important result
\begin{equation}
F_{t+1}^{(32,t)}= F_{t+1}^{(32,242)}+F_{t+1}^{(243,1032)}+\cdot
\cdot \cdot+F_{t+1}^{(59049,99999)}+F_{t+1}^{(100000,t)}.
\end{equation}
Similarly, the initial probability distribution is $f_{i}(i) =
(1,0,0, ... ) = e_{1}$ for any $i$. Thus the same algorithm based
on (20) can be used to compute the degree distribution $P(k, t)$
for this network.

From the computation results, we plot the $\log-\log$ curves for
$P(k, t)$ for some different $m$ and $t$ as shown in Figure 2.
We also list some numerical results in the Table 3. From the
figure and the table, it is clear that this network self-organizes
into a non-stationary scale-free network, with the degree exponent
$\gamma \approx 3.5$.

\begin{table} [htbp]
 \begin{center}
 \caption{ \ \  Numerical results of the power function case}
 \vspace{3pt}
 \begin{tabular} {|c|c|c|c|c|}
 \hline
 parameter $m$ & time $t$ & exponent $\gamma$ & coefficient $c$  \\
 \hline
 1 & 150000 & 3.502938 & 891.641  \\
 3 & 100000 & 3.499978 & 8213.46  \\
 3 & 150000 & 3.502746 & 10920.8  \\
 3 & 200000 & 3.496971 & 12300.2  \\
 5 & 150000 & 3.503176 & 37303.5  \\
 \hline
 \end{tabular}
 \end{center}
 \end{table}

\subsection{Logarithmic function}
\ \ \ \  Let the number of new links in time step $t$ be $m \ln t$.

\ \ \ \ 




We note that after $t$ time steps, the model leads to a random
network with $N=t+m_{0}$ nodes and approximately $\int_{0}^{t}m\ln
xdx$ links. Then, the total degree number of the system at time
$t$ is
\begin{equation}
\sum_{j}k_{j} \approx  2\int_{0}^{t}m\ln xdx = 2mt(\ln t-1).
\end{equation}
The average degree of the system is $\overline{k}=2m(\ln t-1)$,
i.e., it follows a logarithmic law.
There has been no analytical
results for the degree distribution for this case
as, we believe, it is extremely difficult if not impossible.

We now construct the Markov chain for the degree sequence $\{K_{i}(t), t=i,i+1,...\}$.
The state space is $\Omega=\{m_{i},m_{i}+1,...\}$, where $m_{i}=m[\ln i]$.
At time $t$, the probability that an existing node $i$ will connect with
the new node is given by
\begin{equation}
 m\ln t \frac{k_{i}}{\sum_{j}k_{j}} \approx \frac{k_{i}\ln
t} {2t(\ln t-1)}.
\end{equation}
Hence, the one-step transition probabilities are
\begin{equation}
p_{kj}(t+1) = P\{K_{i}(t+1)=j \mid K_{i}(t)=k\}
 = \left\{
 \begin{array}{ll}
 1-\frac{k\ln t}{2t(\ln t-1)}, & j=k \\
 \frac{k\ln t}{2t(\ln t-1)}, & j=k+1 \\
 0, & otherwise
 \end{array}
 \right.
\end{equation}
for $k=m_{i},...,m_{i}+t-i$, and
\begin{equation}
p_{kj}(t+1) = \left\{
 \begin{array}{ll}
 1, & j=k \\
 0, & j\neq k
\end{array}
 \right.
\end{equation}
for $k > m_{i}+t-i$. The transition probability matrix is
\begin{equation}
P_{i}(t+1)=\left[\matrix{1-\frac{m_{i}\ln t}{2t(\ln t-1)} &
\frac{m_{i}\ln t}{2t(\ln t-1)} & \cr
 & \cr
 & 1-\frac{(m_{i}+1)\ln t}{2t(\ln t-1)} & \frac{(m_{i}+1)\ln t}{2t(\ln t-1)} & \cr
 &&\ddots & \ddots & \cr
 & \cr
 &&&1-\frac{(m_{i}+t-i)\ln t}{2t(\ln t-1)} & \frac{(m_{i}+t-i)\ln t}{2t(\ln t-1)} & \cr
 & \cr
 &&&&1 & 0  & \cr
 &&&&&\ddots & \ddots &
 }  \right]
\end{equation}
for $t = i,i+1,...$ .

We note that the structure of the transition matrices here is
similar to that of (12). The difference is that now $m_i$ is not a
constant, in general, but a step function of $i$, as shown in
Table 4.

\begin{table}[htbp]
 \begin{center}
 \caption{ \ \  Intervals of $m_i$ keep constant}
 \vspace{5pt}
 \begin{tabular} {|c|c|c|c|c|c|c|c|c|c|c|}
 \hline
time $i$ & 21 & 55 & 149 & 404 & 1097 & 2981 & 8104 & 22027 & 59875 & 162755 \\
 \hline
$[\ln i]$ & 3 & 4 & 5 & 6 & 7 & 8 & 9 & 10 & 11 & 12 \\
 \hline
 \end{tabular}
 \end{center}
 \end{table}

Therefore, relations (16), (17) and (18) hold for each interval,
e.g., the interval $(404,1096)$. Thus we obtain
the following important result
\begin{equation}
F_{t+1}^{(21,t)}= F_{t+1}^{(21,54)}+F_{t+1}^{(55,148)}+\cdot \cdot
\cdot+F_{t+1}^{(59875,162754)}+F_{t+1}^{(162755,t)}.
\end{equation}
Similarly, the initial probability distribution is $f_{i}(i) =
(1,0,0, ... ) = e_{1}$ for any $i$. Thus the same algorithm based
on (20) can be used to compute the degree distribution $P(k, t)$
for this network.

From the computation results, we plot the $\log-\log$ curves for
$P(k, t)$ for some different $m$ and $t$ as shown in Figure 3.
We also list some numerical results in Table 5. From the figure
and the table, it is clear that this network self-organizes into a
non-stationary scale-free network, with the degree exponent
$\gamma \approx 3.1$ and a positive, though very small,
non-stationary exponent $z$.


\begin{table} [htbp]
 \begin{center}
 \caption{ \ \  Numerical results of the logarithmic function case}
 \vspace{3pt}
 \begin{tabular} {|c|c|c|c|c|}
 \hline
 parameter $m$ & time $t$ & exponent $\gamma$ & coefficient $c$  \\
 \hline
 1 & 150000 & 3.169873 & 542.9149  \\
 3 & 100000 & 3.117526 & 1539.876  \\
 3 & 150000 & 3.081926 & 1722.288  \\
 3 & 200000 & 3.050253 & 1952.588  \\
 5 & 150000 & 3.029171 & 2823.681  \\
 \hline
 \end{tabular}
 \end{center}
 \end{table}

\section{Conclusions and discussions}
\ \ \ \  In summary, we introduce a Markov chain-based new method
to calculate degree distributions of scale-free networks
numerically. Comparing with the existing analytical methods, this
method is more flexible. It offers
the asymptotic property of the degree distribution for the more
complicated models. Using only the transition probability matrix
$P_{i}(t+1)$, we can compute the degree distribution $P(k)$. Since
the complexity of our algorithm is $O(t^2)$, its advantage
over the simulation method is also quite obvious:
it is fast and, for problems that it can handle,
it is more reliable and provides better understanding of
the network behavior.

The use of Markov chain to model the degree evolution is quite
novel and opens the door for the applications of methodologies and
results from a very mature field to the exciting new field of
scale-free networks. For instance, we may consider to compute the
joint degree distribution of a node pair by using Markov chains.
Furthermore, the fact that the evolution of a complex network can
be modeled by Markov chains may indicate an important direction
for us to investigate the underlying mechanisms of growth networks,
since we have accumulated extensive understanding of the structural
properties of Markov chains as we use them to study many natural phenomena.


\newpage

\begin{figure}[htbp]
\center{\psfig {file=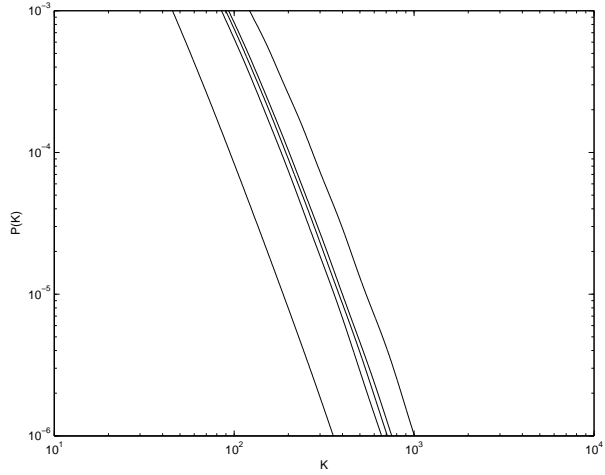, width=8cm}} \caption{{\small The
degree distribution of the power function case }} \label{fig 2}
\end{figure}

In Figure \ref{fig 2}, the five lines from left to right
correspond to three cases: (1) $m=1, ~t=150,000;$ (2)
$m=3,~t=100,000,~ 150,000,~200,000;$ (3) $m=5, ~t=150,000.$ In
(2), the three lines are separated, demonstrating the
non-stationarity of the degree distribution. Again, we can see
that the degree exponents are essentially independent of $m$ as
the lines are parallel to each other.

\bigskip
\bigskip

\begin{figure}[htbp]
 \begin{center}
  \includegraphics[width=8cm]{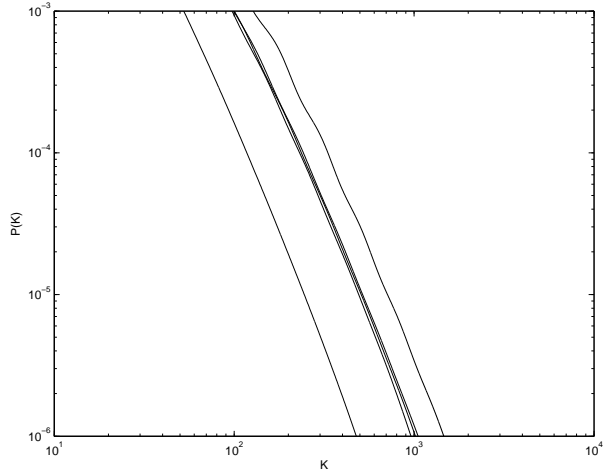}\\
 \caption{{\small The degree distribution of the
logarithmic function case}} \label{fig 3}
\end{center}
\end{figure}

In Figure \ref{fig 3}, the five lines from left to right
correspond to three cases: (1) $m=1, ~t=150,000;$ (2)
$m=3,~t=100,000,~ 150,000,~200,000;$ (3) $m=5, ~t=150,000.$ In
(2), the three lines are very close to each other but not entirely
overlapping, showing that while the degree distribution is not
stationary, the non-stationary exponent is very small. Again, we
can see that the degree exponents are essentially independent of
$m$ as the lines are parallel to each other.

\end{document}